\documentclass[%
 aip,
 amsmath,amssymb,
 reprint,%
]{revtex4-1}

\usepackage{graphicx}
\usepackage{dcolumn}
\usepackage{bm}

\usepackage[utf8]{inputenc}
\usepackage[T1]{fontenc}
\usepackage{mathptmx}
\usepackage{etoolbox}
\usepackage[section]{placeins}
\usepackage{comment}
\usepackage{xfrac}

\makeatletter
\def\@email#1#2{%
 \endgroup
 \patchcmd{\titleblock@produce}
  {\frontmatter@RRAPformat}
  {\frontmatter@RRAPformat{\produce@RRAP{*#1\href{mailto:#2}{#2}}}\frontmatter@RRAPformat}
  {}{}
}%

\begin{document}

\preprint{AIP/123-QED}

\title[Applied Surface Science]{A Two-Step Chemical Vapor Deposition Process for the Synthesis of an Ir(111)/Borophene/2D-Hexagonal Boron Nitride Heterostructure by Intrinsic Segregation}

\author{Marko A. Kriegel} 
\affiliation{Faculty of Physics, University of Duisburg-Essen, Lotharstr. 1, 47057 Duisburg, Germany}%

\author{Karim M. Omambac}
\altaffiliation[present address: ]{Department of Engineering Physics, Ecole Polytechnique de Montreal, C.P. 6079, Succ. Centre-Ville, Montreal QC H3C~3A7, Canada}
\affiliation{Faculty of Physics, University of Duisburg-Essen, Lotharstr. 1, 47057 Duisburg, Germany}

\author{Smruti R. Mohanty}
\affiliation{Faculty of Physics, University of Duisburg-Essen, Lotharstr. 1, 47057 Duisburg, Germany}

\author{Tobias Hartl}
\affiliation{Institute of Physics II, University of Cologne, Z\"{u}lpicher Str. 77, 50937 Cologne, Germany}

\author{Stefan Schulte}
\altaffiliation[present address: ]{Peter Gru\"{u}nberg Institut (PGI-3), Forschungszentrum J\"{u}lich, 52425 J\"{u}lich, Germany}
\affiliation{Institute of Physics II, University of Cologne, Z\"{u}lpicher Str. 77, 50937 Cologne, Germany}

\author{Niels Ganser}
\affiliation{Faculty of Physics, University of Duisburg-Essen, Lotharstr. 1, 47057 Duisburg, Germany}%

\author{Pascal Dreher}
\altaffiliation[present address: ]{Institute for Physical and Theoretical Chemistry, University of W\"{u}rzburg, Am Hubland S\"{u}d, 97074 W\"{u}rzburg, Germany}
\affiliation{Faculty of Physics, University of Duisburg-Essen, Lotharstr. 1, 47057 Duisburg, Germany}%

\author{Alexandra R\"{o}dl}
\affiliation{Faculty of Physics, University of Duisburg-Essen, Lotharstr. 1, 47057 Duisburg, Germany}%

\author{Steffen Franzka}
\affiliation{Interdisciplinary Center for Analytics on the Nanoscale (ICAN), Carl-Benz Str. 199, 47057 Duisburg, Germany}

\author{Germ\'an Sciaini}
\affiliation{Ultrafast Electron Imaging Lab, Department of Chemistry, University of Waterloo, 200 University Ave. West, Waterloo ON N2L~3G1, Canada}

\author{Thomas Michely}
\affiliation{Institute of Physics II, University of Cologne, Z\"{u}lpicher Str. 77, 50937 Cologne, Germany}

\author{Frank-J. Meyer zu Heringdorf}
\affiliation{Faculty of Physics, University of Duisburg-Essen, Lotharstr. 1, 47057 Duisburg, Germany}%
\affiliation{Interdisciplinary Center for Analytics on the Nanoscale (ICAN), Carl-Benz Str. 199, 47057 Duisburg, Germany}
\affiliation{Center for Nanointegration Duisburg-Essen (CENIDE), Carl-Benz-Str. 199, 47057 Duisburg, Germany}%

\author{Michael Horn-von Hoegen}
\affiliation{Faculty of Physics, University of Duisburg-Essen, Lotharstr. 1, 47057 Duisburg, Germany}%
\affiliation{Center for Nanointegration Duisburg-Essen (CENIDE), Carl-Benz-Str. 199, 47057 Duisburg, Germany}%

\email{mhvh@uni-due.de}

\date{\today}

\begin{abstract}
We report on a two-step ultrahigh vacuum chemical vapor deposition synthesis of a vertical Ir(111)/borophene/hexagonal boron nitride heterostructure, using borazine as a single-source precursor. The process takes advantage of the finite solubility of boron in Ir: low precursor pressure at high temperature first establishes a boron reservoir in the near-surface region of the substrate, whereas subsequent growth at higher precursor pressure promotes the formation of a closed hexagonal boron nitride monolayer. During cooldown, the reduced boron solubility drives segregation to the surface, resulting in the formation of a borophene monolayer beneath the hexagonal boron nitride overlayer. The heterostructure, with micron sized grains, homogeneously covers the entire Ir substrate. The study is performed by complementary spot profile analysis low-energy electron diffraction, low-energy electron microscopy, and scanning tunneling microscopy measurements. This \textit{intrinsic segregation}-assisted growth concept provides a promising route toward scalable synthesis of high-quality, vertical heterostructures of two-dimensional materials.\\


\end{abstract}

\maketitle

\section*{\label{sec:level3}Introduction}

Since the discovery of two-dimensional materials (2DMs)~\cite{Novoselov.2004, Geim.2007} vertically stacked heterostructures have been regarded as one of the most promising routes toward tailored material functionalities~\cite{Geim.2013}. By combining atomically thin layers with semiconducting, insulating, or metallic properties, such van der Waals heterostructures provide artificial material systems with functionalities that are absent in the individual constituents~\cite{Schwierz.2022}. As a result of that, the controlled fabrication of such heterostructures has become a central goal of 2DM research~\cite{Liu.2016, Ponomarenko.2011,Qi.2023}.

So far, the scientifically most successful approach has been the deterministic dry transfer and sequential stacking of mechanically exfoliated 2DM layers\cite{Geim.2013,Fan.2024,Onodera.2024}. While this strategy has enabled the realization of a wide range of model systems and numerous technological breakthroughs~\cite{Britnell.2013,Furchi.2014,Withers.2015}, it has also revealed the rich physics of van der Waals heterostructures~\cite{Behura.2021,Shah.2024,Algarni.2025}. At the same time, however, it is inherently limited in terms of scalability, reproducibility, and compatibility with larger-scale processing. In contrast, direct growth methods, such as chemical vapor deposition (CVD)\cite{Sun.2021,Zhang.2013}, are much better suited for scalable synthesis for technological applications. However, they usually favor the formation of only a single 2DM layer\cite{Zhao.2013}, when performed in ultrahigh vacuum (UHV). Once a closed 2DM layer has formed, catalytic decomposition of the precursor of another 2DM is strongly suppressed, which prevents straightforward growth of an additional 2DM on top. Not surprisingly, the direct synthesis of vertical heterostructures remains a major challenge. Only in a limited number of cases has this bottleneck been overcome. Strategies include the use of multiple precursors~\cite{Cuxart.2021} or the combination of different growth and deposition techniques\cite{Qi.2018,CastellanosGomez.2022,Bhakta.2025}, such as plasma-enhanced chemical vapor deposition (PECVD)\cite{Sterling.1965,Adhikari.2016,Seok.2021}, metal-organic chemical vapor deposition (MOCVD)\cite{Manasevit.1969,Kang.2015}, or molecular beam epitaxy (MBE)\cite{Ugeda.2014,Singh.2022}.

Here, we demonstrate a route to a vertical borophene/\textit{h}BN heterostructure on Ir(111), formed during CVD, by exploiting the \emph{intrinsic segregation} of boron after growth of an \textit{h}BN monolayer. We use Ir(111) as a model system to understand how the catalytic processes work on a well-studied substrate. Using borazine as a single-source precursor under ultrahigh vacuum (UHV) conditions, elemental boron released during precursor dissociation was first dissolved into the Ir substrate at high sample temperature. Subsequently, by strongly increasing the borazine dosing pressure $p_\text{dose}$ and reducing the growth temperature $T_\text{g}$, the chemical balance was shifted toward the growth of a closed monolayer of \textit{h}BN. Upon further cooling, the decreasing boron solubility in Ir induces segregation to the surface~\cite{Omambac.2021}, and thereby the formation of a complete borophene layer beneath the \textit{h}BN top layer.

\section*{Experimental}

The growth process and the resulting vertical heterostructure were characterized by spot-profile analysis low-energy electron diffraction (SPA-LEED)~\cite{MHvH.1999}, low-energy electron microscopy (LEEM), and scanning tunneling microscopy (STM)  under UHV conditions at a base pressure of $2 \times 10^{-10}$\,mbar. Borophene and \textit{h}BN were synthesized on a well-oriented single-crystalline Ir(111) substrate (Mateck, 99.99\,\%). Prior to growth, the Ir crystal was cleaned by repeated cycles of Ar$^+$ sputtering, annealing, and oxygen etching. For SPA-LEED experiments, the cleaning procedure was continued until the full width at half maximum (FWHM) of the diffraction spots reached the instrumental resolution limit, corresponding to approximately 0.5\,\% of the surface Brillouin zone. For the other methods, surface preparation was considered complete once a clean surface with a low density of Ir steps was observed in phase-contrast LEEM and STM.

LEEM experiments were performed using an ELMITEC SPE-LEEM III instrument at the University of Duisburg-Essen. This UHV microscope enables real-time imaging during growth, even at temperatures of more than 1200\,$^\circ$C, as well as throughout the subsequent cooling process, with frame rates of up to 4.5 frames per second. The microscope was operated in both imaging and diffraction modes.

On Ir(111), \textit{h}BN, borophene, and the heterostructure were prepared by catalytic decomposition of borazine (B$_3$N$_3$H$_6$) at partial pressures in the range from $1 \times 10^{-7}$ to $1 \times 10^{-6}$\,mbar and substrate temperatures between 890\,$^\circ$C and 1140\,$^\circ$C. Heating was achieved by electron bombardment from the backside of the sample. Temperatures were measured by a W-Re-thermocouple that was spot welded to the sample holder. To prevent unintended precursor decomposition, borazine was kept at temperatures $\leq$\,-5\,$^\circ$C in a home-built Peltier-cooled reservoir at all times. Before synthesis, the gas line was evacuated with a turbomolecular pump to remove residual contaminants. SPA-LEED patterns were recorded at room temperature at an electron energy of 69\,eV. 

For STM investigations, the Ir(111) crystal was cleaned by repeated cycles of 2\,keV Xe$^+$ sputtering and flash annealing to 1200\,$^\circ$C. STM measurements were performed at room temperature using tunneling currents between 0.2\,nA and 4\,nA and sample bias voltages ranging from $-3$\,V to $+3$\,V.

\section*{Results and Discussion} 
\subsection*{SPA-LEED}
\begin{figure*}[htbp]
    \centering
    \includegraphics[width=1\linewidth]{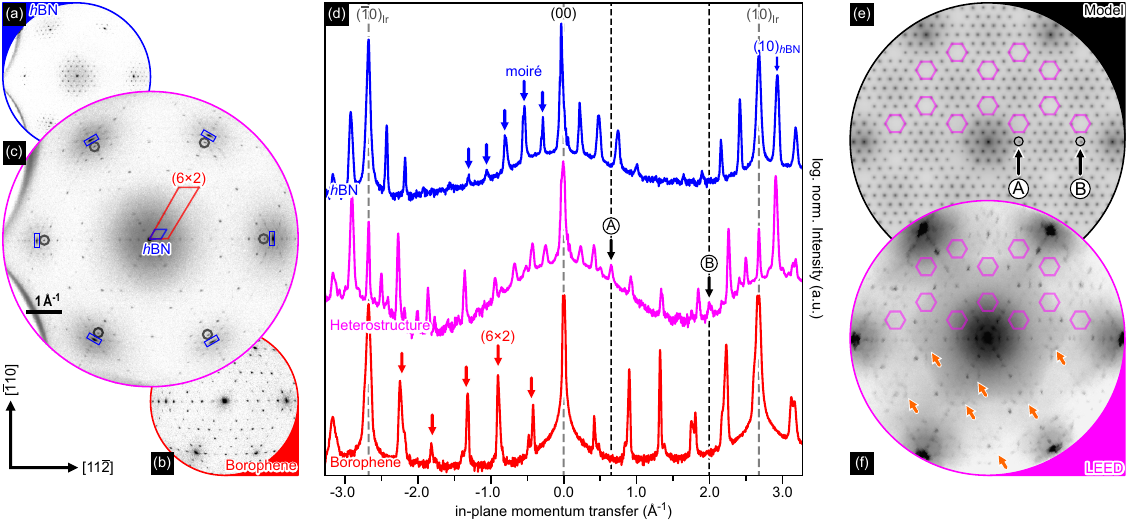}
    \caption{SPA-LEED patterns in inverse logarithmic intensity scale taken at 69\,eV from (a) an \textit{h}BN layer with moir\'e spots surrounding all integer order spots, (b) a boro phene layer with $(6\times2)$ reconstruction, and (c) the heterostructure with the first integer order spots of borophene and \textit{h}BN indicated by circles and rectangles, respectively. The unit cells of \textit{h}BN moiré pattern and borophene are indicated by blue and red parallelograms. (d) LEED spot profiles along the [$11\bar{2}$] direction for \textit{h}BN (top) heterostructure (center) and borophene (bottom) in a logarithmic intensity scale. Integer order spots are indicated by vertical gray dashed lines. The positions of moir\'e spots and 6-fold spots for \textit{h}BN and borophene are indicated on the left hand side by vertical arrows in blue and red, respectively. Dashed black lines on the right hand side indicate spots for the heterostructure which are absent in the two reference structures. (e) Modeled diffraction pattern through convolution of a $(6\times2)$ reciprocal lattice with a first-order moir\'e reciprocal lattice. The spot-free regions, so called hollow hexagons, are indicated in blue and reproduce the experimental findings. (f) LEED pattern taken in LEEM at 50\,eV for comparison. Hollow hexagons are indicated in pink. Orange arrows indicate $(6\times2)$ spots surrounded by moir\'e spots.}
    \label{fig: Only SPA-LEED}
\end{figure*}

Prior to analyzing the heterostructure (HET) in SPA-LEED, we prepared single-layer \textit{h}BN and borophene on Ir(111) as reference systems. The \textit{h}BN layer was grown by self-limiting CVD at $T_\mathrm{g} = 960\,^\circ$C and a borazine pressure of $p_\mathrm{dose} = 1 \times 10^{-6}$\,mbar~\cite{Petrovic.2017,Omambac.2023}. Its LEED pattern, shown in Fig.~\ref{fig: Only SPA-LEED}(a), exhibits a dense hexagonal array of moir\'e spots originating from the incommensurate near $23:21$ lattice matching of the $R0$-oriented \textit{h}BN overlayer with the Ir(111) substrate~\cite{Kriegel.2023}. The periodic change of registry between the \textit{h}BN and the underlying substrate causes an undulation of the 2DM, in turn producing the satellite spots of the moir\'e pattern~\cite{Horn-vonHoegen.1993}.

In contrast, borophene was prepared by a ``growth from below'' approach, as described elsewhere for a variety of metallic substrates\cite{Omambac.2021, Kiraly.2019, Sutter.2021}. Using a significantly higher temperature, $T_\mathrm{g} =1140\,^{\circ}$C, and lower precursor pressure, $p_\mathrm{dose} = 1 \times 10^{-7}$\,mbar, the borazine dissociates into its elemental constituents; nitrogen desorbs, while boron dissolves into the near-surface region of the Ir substrate. During subsequent cooling below $\sim 890\,^\circ$C, the reduced boron solubility induces segregation to the surface and thereby the formation of borophene.

As shown in Fig.~\ref{fig: Only SPA-LEED}(b), the resulting LEED pattern is governed by the commensurate $(6 \times 2)$ periodicity of the $\chi_6$ borophene phase\cite{Omambac.2021,Radatovic.2022,Vinogradov.2019}. Because three rotational domains coexist on the threefold-symmetric substrate, their incoherent superposition produces an apparently hexagonal diffraction pattern. In particular, the absence of diffraction spots at $(\sfrac{1}{6},\sfrac{1}{6})$, $(\sfrac{2}{6},\sfrac{2}{6})$, $(\sfrac{4}{6},\sfrac{1}{6})$, $(\sfrac{1}{6},\sfrac{4}{6})$, and symmetry equivalent ones is indicative of the incoherent superposition of three $(6 \times 2)$ patterns rotated by $120^\circ$ relative to each other.

The vertical heterostructure was synthesized by a combination of the growth procedures of both the \textit{h}BN and borophene (see LEEM results). The LEED pattern of the heterostructure, shown in Fig.~\ref{fig: Only SPA-LEED}(c), exhibits a significantly richer set of diffraction features than the two reference structures. The first-order diffraction spots of the Ir(111) substrate, located at $k = 2.67$\,\AA$^{-1}$ and marked by dark circles, are weaker than in the reference patterns. This weakened Ir spot intensity indicates the formation of the two-layer vertical heterostructure on top of the Ir surface. In contrast, the first-order spots of the \textit{h}BN overlayer at $k = 2.92$\,\AA$^{-1}$, indicated by blue rectangles, are more intense and slightly elongated along the azimuthal direction.

To assign these diffraction features, the diffraction spot positions of the heterostructure are compared with those of the \textit{h}BN and borophene reference lattices. The corresponding intensity line profiles are displayed in Fig.~\ref{fig: Only SPA-LEED}(d), showing the \textit{h}BN moir\'e pattern (blue), the heterostructure (pink), and the borophene $(6 \times 2)$ pattern (red), from top to bottom. Dashed gray lines indicate the positions of the specular $(00)$ spot and the first-order diffraction spots of Ir(111) and \textit{h}BN, whereas the moir\'e and $(6 \times 2)$ spot positions are marked by vertical blue and red arrows, respectively. Both, the \textit{h}BN and heterostructure line profiles exhibit the so-called bell-shaped component, i.e., very broad diffuse intensity with a FWHM on the order of 50\,\% of the Brillouin zone, underlying all integer-order diffraction spots. This pronounced feature is found for many weakly bonded two-dimensional material systems~\cite{Chen.2019,Chen.2020,Omambac.2021b,Petrovic.2021}. In contrast, it is absent in the bottom profile, consistent with the much stronger binding of borophene to the Ir substrate~\cite{Vinogradov.2019,Ugolotti.2025}.

Many diffraction spots of the heterostructure coincide with spots already present in either the \textit{h}BN or borophene reference pattern. However, additional spots appear that cannot be accounted for by a simple incoherent superposition of the two patterns. Two such spots, with their positions indicated by vertical dashed black lines, marked A and B, and by arrows on the right-hand side in Fig.~\ref{fig: Only SPA-LEED}(d) of the line profiles, are observed only for the heterostructure. Their positions at $0.70$\,\AA$^{-1}$ and $1.98$\,\AA$^{-1}$ are naturally explained by a convolution of the two reciprocal lattices: Combining the $(6 \times 2)$ reciprocal lattice vector, $k_{(6 \times 2)} = 0.445$\,\AA$^{-1}$, with the \textit{h}BN moir\'e wave vector, $k_\mathrm{moir\acute{e}} = 0.254$\,\AA$^{-1}$, yields $k_\mathrm{hetero} = 0.70$\,\AA$^{-1}$ and $1.97$\,\AA$^{-1}$, in agreement with the experimental finding. This convolution in reciprocal space is equivalent to a multiplication of the real space scattering amplitudes of the two layers, and thus shows that the incident electrons are diffracted coherently by both the \textit{h}BN and the borophene layer.

To further support the hypothesis of vertical stacking and coherent diffraction, the diffraction pattern was modeled by convoluting the $(6 \times 2)$ borophene pattern (with its three rotational domains) with the \textit{h}BN moir\'e lattice, as shown in Fig.~\ref{fig: Only SPA-LEED}(e). In this procedure, only the six first-order moir\'e spots at $k_\mathrm{moir\acute{e}} = \pm 0.254$\,\AA$^{-1}$ were considered. The resulting model pattern reproduces the manifold of diffraction spots observed experimentally in Fig.~\ref{fig: Only SPA-LEED}(c). In particular, the two spots only found in the heterostructure linescan can also be reconstructed by the model and are indicated by black circles and arrows. Another particularly distinctive feature is the presence of spot-free regions, highlighted by pink hexagons in Fig.~\ref{fig: Only SPA-LEED}(e), which are also clearly visible in Fig.~\ref{fig: Only SPA-LEED}(f), showing a LEED image recorded at 50\,eV in LEEM.

\begin{figure}[htbp]
    \centering
    \includegraphics[width=\columnwidth]{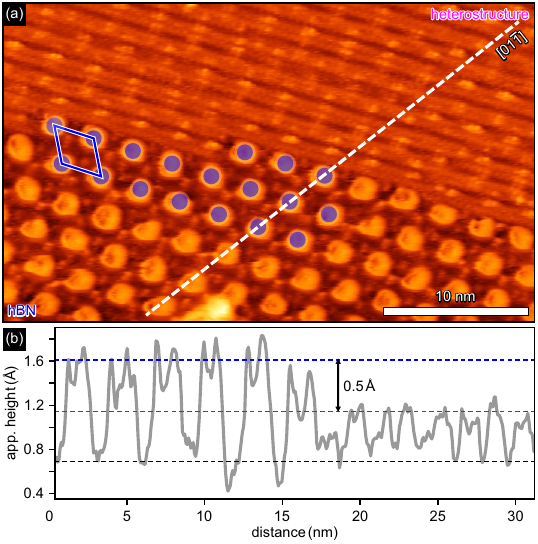}
    \caption{(a) Constant-current STM image recorded at $U_\mathrm{bias}=2.0$\,V and $I_\mathrm{t}=3.7$\,nA with a field of view of $22 \times 36$\,nm$^2$. The lower part of the image shows \textit{h}BN on Ir(111) with the characteristic pore and wire structure of the hexagonal moir\'e structure. The upper part reveals the borophene/\textit{h}BN heterostructure, characterized by stripe-like corrugations with a spacing of 1.4\,nm, consistent with the $(6 \times 2)$ borophene structure. The \textit{h}BN moir\'e modulation is only faintly visible in the heterostructure region. (b) Apparent height profile extracted along the [01$\bar{1}$] direction and indicated by the dashed line in (a). The left section corresponds to \textit{h}BN and exhibits the moir\'e periodicity of $\sim$29\,\AA\cite{FarwickzumHagen.2016}, while the right section corresponds to the heterostructure and shows the $6a_\mathrm{Ir}$ periodicity. The reduced apparent height of the heterostructure relative to the \textit{h}BN region is attributed to the insulating nature of the electronically decoupled \textit{h}BN top layer\cite{Cuxart.2021}.}
    \label{fig: STM}
\end{figure}

\subsection*{STM}
The weaker moir\'e spots of the heterostructure can be explained by a weaker interaction between \textit{h}BN and borophene, as reported by Cuxart \textit{et~al.}~\cite{Cuxart.2021}. We therefore conducted STM measurements on a partially formed heterostructure, prepared by the same procedure as used in the LEED experiments. In this case, however, the duration of the initial boron loading of the Ir substrate was reduced. This lead to a partial coverage with borophene and enabled to simultaneously image both, the \textit{h}BN reference phase and the heterostructure, as shown in Fig.~\ref{fig: STM}(a). The STM image was recorded at room temperature in constant-current mode with a positive sample bias of $U_\mathrm{bias} = 2.0$\,V and a tunneling current of $I_\mathrm{t} = 3.7$\,nA.
Pore and wire like features of the \textit{h}BN moir\'e structure are clearly resolved in the lower part of the image~\cite{Corso.2004,Berner.2007,Goriachko.2007,FarwickzumHagen.2016,Will.2018}. These pronounced real space modulations (with the moir\'e unit mesh indicated as a blue rhombus) cause the manifold of moir\'e spots observed in LEED. Unlike the \textit{h}BN reference area, the heterostructure region in the upper part of the image appears as a stripe-like phase with only a weak additional height modulation. Here, the measured spacing between adjacent stripes is $\sim 1.4$\,nm, corresponding to the lateral sixfold periodicity of the $(6 \times 2)$ reconstruction of borophene. The weak superimposed height modulation is not observed for bare borophene and is therefore attributed to the vertical stacking with the \textit{h}BN layer~\cite{Omambac.2021}.

\begin{figure}[htbp]
    \centering
    \includegraphics[width=\columnwidth]{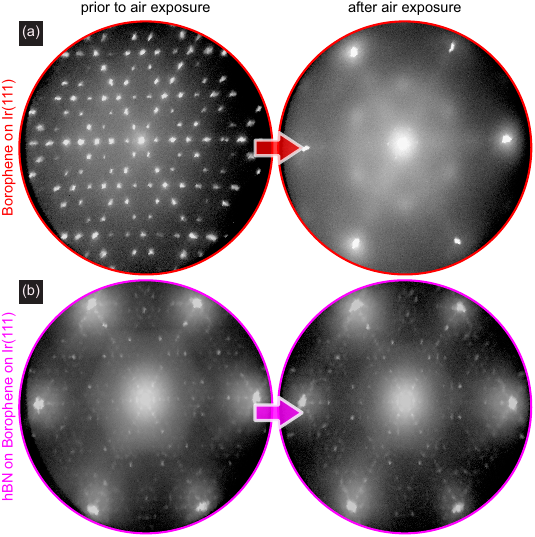}
    \caption{LEED patterns taken in LEEM for (a) a borophene reference layer and (b) the heterostructure, prior to (left column) and after (right column) exposure to ambient conditions. The characteristic $(6 \times 2)$ diffraction pattern of borophene disappears completely, and only weak, diffuse intensity remains, indicating oxidation of the boron layer. The diffraction pattern of the heterostructure, however, remains essentially unchanged upon air exposure.}
    \label{fig: Exposure}
\end{figure}

The weak, \textit{h}BN-related features within the heterostructure region, together with the reduced apparent height of this phase [see Fig.~\ref{fig: STM}(b)], reflect the modified electronic properties of the \textit{h}BN layer in the heterostructure. The \textit{h}BN contributes only weakly to the tunneling current, consistent with its insulating character which makes \textit{h}BN mostly transparent to the STM~\cite{Will.2018}. This observation suggests that the \textit{h}BN layer is electronically decoupled from the Ir substrate by the borophene layer and therefore forms the top layer of this vertical heterostructure. Consequently, the weak interaction allows for rotational disorder of the \textit{h}BN domains, which causes the broadening of moir\'{e} spots, as observed in the LEED measurements shown in Fig.~\ref{fig: Only SPA-LEED}(c). Such behavior was also reported by the Auw\"{a}rter Group~\cite{Cuxart.2021}.

\begin{figure*}[htb]
    \centering
    \includegraphics[width=1\linewidth]{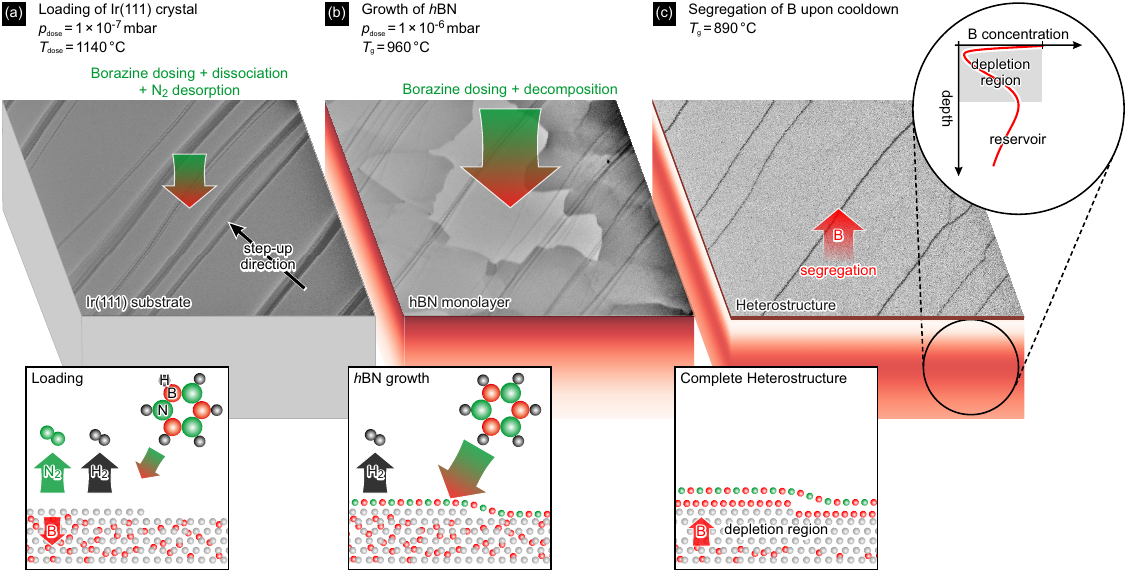}
    \caption{Bright-field LEEM images acquired at 3.9\,eV electron energy \emph{in-situ} at elevated temperatures during the formation of the heterostructure. Field of view is  $15 \times 15$\,\textmu m$^2$. (a) Exposure of the Ir(111) substrate to borazine at $T_\mathrm{dose}=1140\,^\circ$C and $p_\mathrm{dose}=1 \times 10^{-7}$\,mbar until a complete \textit{h}BN monolayer was formed. Under these conditions, the precursor dissociates into its elemental constituents: nitrogen and hydrogen desorb, whereas boron dissolves into the Ir substrate. The faint dark lines are single atomic steps of the Ir surface. (b) Self-limited growth of an \textit{h}BN monolayer at $T_\mathrm{g}=960\,^\circ$C and $p_\mathrm{dose}=1 \times 10^{-6}$\,mbar. The weak contrast variations indicate the formation of twin domains\cite{Petrovic.2017}, while the substrate step structure remains visible. The red color gradient sketches the Boron concentration in the bulk. (c) Formation of the heterostructure during cooldown at $T_\mathrm{HET}\approx 890\,^\circ$C by boron segregation to the surface. Borophene-induced step bunching leads to a pronounced change in the surface step morphology. The circular inset depicts the boron concentration and its depletion in the subsurface region in accordance with TOF-SIMS results published in~\cite{Omambac.2021}. 
    The sketches in the square boxes illustrate the key processes of precursor dissociation, desorption, dissolution, growth, and segregation during the formation of the heterostructure.}
    \label{fig: Segregation}
\end{figure*}

\subsection*{LEED \& LEEM}
To verify our hypothesis regarding the stacking sequence, we made use of the well-established chemical inertness of \textit{h}BN.
Therefore, \textit{h}BN is commonly used as a protective capping layer for sensitive two-dimensional materials and nanostructures against oxidation~\cite{Sirota.2018,BartusPravda.2022}. 
In contrast, borophene oxidizes rapidly under exposure to oxygen or ambient conditions \cite{Omambac.2021}. 

We exposed both, a reference borophene layer and the heterostructure to air for 8~min. The corresponding LEED patterns recorded before (left column) and after exposure (right column) are shown in Fig.~\ref{fig: Exposure}.
As expected, the bare borophene layer exhibits a pronounced modification after exposure to ambient conditions. The characteristic $(6 \times 2)$ diffraction spots disappear completely. Instead, only weak and diffuse intensity remains in the region between the specular $(00)$ spot and the first-order diffraction spots of the Ir substrate, clearly indicating oxidation of the borophene layer.
In striking contrast, the diffraction pattern of the heterostructure remains essentially unchanged after exposure to ambient conditions. This result demonstrates that the top \textit{h}BN layer effectively protects the underlying borophene against oxidation. Thus, we conclude that borophene is located between the Ir substrate and the \textit{h}BN layer above. This interpretation is consistent with previous work from Auw\"arter and co-workers~\cite{Cuxart.2021}.

Finally, LEEM was employed to investigate the spatial homogeneity of the heterostructure formation. Figure~\ref{fig: Segregation} presents a sequence of bright-field LEEM images recorded before, during, and after formation of the heterostructure from the same growth experiment. All images were recorded at elevated temperature during growth and cover a field of view of $15 \times 15$\,\textmu m$^2$.

Figure~\ref{fig: Segregation}(a) shows the Ir(111) surface during the initial boron loading step at a borazine pressure of $1 \times 10^{-7}$\,mbar and a growth temperature of $T_\mathrm{g} = 1140\,^\circ$C. The faint horizontal dark lines correspond to single atomic steps with a height of $2.22$\,\AA~\cite{Coraux.2008}. At this stage, no change in the diffraction pattern is observed, consistent with adsorption of the precursor and dissociation into its elemental constituents: nitrogen and hydrogen desorb as molecules~\cite{Lu.2021,Haug.2020}, whereas boron dissolves into the Ir bulk~\cite{Omambac.2021}.

Upon increasing the dosing pressure by one order of magnitude to $p_\mathrm{dose} = 1 \times 10^{-6}$\,mbar and lowering the growth temperature to $T_\mathrm{g} = 960\,^\circ$C, the chemical balance shifts toward formation of an \textit{h}BN layer~\cite{Omambac.2023}. It is crucial to keep the temperature above $\sim 890\,^\circ$C to prevent the segregation of the boron in the subsurface region~\cite{Omambac.2021}. Figure~\ref{fig: Segregation}(b) shows the surface after completion of a closed \textit{h}BN monolayer in the regime of self-limited growth~\cite{Zhao.2013}. The weak contrast variations are attributed to the presence of twin domains of a size of a few micrometers, which exhibit slightly different electron reflectivities~\cite{Petrovic.2017}. The small bright features are assigned to defect-rich intersections of domain boundaries~\cite{Jong.2023}. Single atomic steps are still visible. The red color gradient beneath the Ir(111) surface displays the boron concentration in the bulk.

During subsequent cooling to approximately $890\,^\circ$C, dissolved boron segregates back to the surface as described by Omambac \textit{et~al.}~\cite{Omambac.2021}. In the present case, it forms a borophene layer underneath the \textit{h}BN overlayer. If boron atoms are dissolved deeper in the bulk than their diffusion length, their segregation is suppressed, resulting in the formation of a boron-depleted region directly beneath the surface~\cite{Omambac.2021,Pan.2021}.The homogeneous contrast in Fig.~\ref{fig: Segregation}(c) indicates that the resulting Ir(111)/borophene/\textit{h}BN heterostructure extends laterally with high spatial uniformity, covering the entire Ir substrate's surface. In addition, the step morphology markedly changes from the single-atomic-step pattern seen in Fig.~\ref{fig: Segregation}(a,b) to step bunches in Fig.~\ref{fig: Segregation}(c). This pronounced restructuring of the Ir surface provides additional evidence for borophene formation beneath the \textit{h}BN top layer, consistent with the step-bunching previously reported for borophene on Ir(111)~\cite{Omambac.2021}.

It should be noted that, experimentally, the heterostructure formation can be achieved either by increasing $p_\mathrm{dose}$ or by decreasing $T_\mathrm{g}$, provided that the phase boundary, separating borophene formation from \textit{h}BN growth in the phase diagram shown in Fig.~4(b) of Ref.~\citenum{Omambac.2023}, is crossed.

\section*{Conclusions}

We have established a synthesis route to a vertical borophene/\textit{h}BN heterostructure that fully covers the surface of the Ir(111) substrate. Using borazine as a single precursor, the heterostructure is formed in a two-step ultrahigh vacuum CVD process controlled by the substrate temperature and the precursor dosing pressure. The process exploits the increased solubility of boron in Ir at elevated temperatures: Under low precursor pressure, boron released during borazine dissociation dissolves into the near-surface region of the substrate. A subsequent increase in precursor pressure and/or reduction of the growth temperature leads to the formation of a closed \textit{h}BN monolayer. Upon cooling, the reduced boron solubility intrinsically drives the boron segregation back to the surface, thereby inducing the formation of borophene beneath the \textit{h}BN layer. We refer to this process as \emph{intrinsic segregation}.

The structural characterization by SPA-LEED confirms the vertical stacking of the two structurally unaffected two-dimensional materials, while the STM study indicates the characteristic ease of interaction between \textit{h}BN and its support. 
In contrast to borophene, the heterostructure remains structurally unchanged upon exposure to ambient conditions. This finding shows that the \textit{h}BN overlayer effectively protects borophene against oxidation, thereby confirming the stacking sequence Ir(111)/borophene/\textit{h}BN. Real-time LEEM further shows that the heterostructure forms homogeneously across the entire surface.

Overall, the present work establishes an intrinsic segregation-assisted growth concept for the fabrication of high-quality borophene/\textit{h}BN heterostructures on Ir(111). Beyond the specific material combination studied here, this approach provides a promising strategy for the scalable synthesis of vertically stacked two-dimensional material systems.

\section*{Author Contributions}
M.K., K.M.O., T.H., St.S., S.F., S.R.M., F.-J.MzH., A.R., and P.D. performed the experiments, supporting measurements, and analyzed the data.
M.K., S.R.M. and N.G. prepared the figures. 
M.HvH. and F.-J.MzH. conceived and supervised the project. 
All authors discussed the results.
The manuscript was written through contributions of M.K., K.M.O., S.R.M., N.G., T.M. and M.HvH.
All authors have given approval to the final version of the manuscript.

\section*{Declaration of Competing Interest}
The authors declare that they have no known competing financial
interests or personal relationships that could have appeared to influence the work reported in this paper.

\section*{Data availability}
Data will be made available on request.

\section*{Acknowledgements}
This work was funded by the Deutsche Forschungsgemeinschaft (DFG, German Research Foundation) – IRTG~2803 – 461605777 and through project B06 of CRC~1242 – 278162697. G.S. acknowledges the support of the Natural Sciences and Engineering Council of Canada (NSERC), [CREATE 565360]


\section*{References}
\bibliography{BibTex_Heterostructure_Paper}

\end{document}